\documentclass[journal]{IEEEtran}

\usepackage{cite}
\usepackage[dvips]{graphicx}
\usepackage[cmex10]{amsmath}
\usepackage{amssymb}
\usepackage{url}

\newcommand{\Ref}[1]{Ref.~\citen{#1}}
\newcommand{\Fig}[1]{Fig.~\ref{#1}}
\newcommand{\Sec}[1]{Sec.~\ref{#1}}
\newcommand{\Tab}[1]{Table~\ref{#1}}

\newcommand{\up}[1]{\ensuremath{^{\mbox{\scriptsize{#1}}}}}
\newcommand{\dn}[1]{\ensuremath{_{\mbox{\scriptsize{#1}}}}}
\newcommand{\EE}[2]{\ensuremath{\mbox{10}^{#1\mbox{\scriptsize{#2}}}}}
\newcommand{\EN}[3]{\ensuremath{
\mbox{#1}\times\mbox{10}^{#2\mbox{\scriptsize{#3}}}}}
\newcommand{\EA}[5]{\ensuremath{
\mbox{#1}^{+\mbox{\scriptsize{#2}}}_{-\mbox{\scriptsize{#3}}}
\times\mbox{10}^{#4\mbox{\scriptsize{#5}}}}}

\newcommand{\Pmul}{\ensuremath{P(\mbox{1}e)}}
\newcommand{\etag}{\ensuremath{\varepsilon(\mbox{tag}|\mbox{1}e)}}

\begin{document}

\title{A Prototype Large-Angle Photon Veto Detector\\ for the P326 Experiment 
at CERN}
\urldef{\moulson}\url{moulson@lnf.infn.it}
\author{F.~Ambrosino, A.~Antonelli, E.~Capitolo, P.S.~Cooper, R.~Fantechi,
L.~Iannotti, G.~Lamanna, E.~Leonardi, M.~Moulson$^*$, 
M.~Napolitano, V.~Palladino,
M.~Raggi, A.~Romano, G.~Saracino, M.~Serra, T.~Spadaro,\\ P.~Valente,  
and~S.~Venditti%
\thanks{Manuscript received November 21, 2007.}%
\thanks{F.~Ambrosino, M.~Napolitano, V. Palladino, A.~Romano, and G.~Saracino
are with the Dipartimento di Scienze Fisiche dell'Universit\`a 
and Sezione INFN, Napoli, Italy.}%
\thanks{A.~Antonelli, E.~Capitolo, P.S.~Cooper, L.~Iannotti, M.~Moulson, 
M.~Raggi, and T.~Spadaro are with the Laboratori Nazionali di Frascati 
dell'INFN, Frascati, Italy.
P.S.~Cooper is a visitor from the Fermi National Accelerator
Laboratory, Batavia IL, USA}%
\thanks{R. Fantechi, G.~Lamanna, and S.~Venditti are with the
Dipartimento di Fisica dell'Universit\`a and Sezione INFN, Pisa, Italy.}%
\thanks{E.~Leonardi, M.~Serra, and P.~Valente are with the Dipartimento di
Fisica dell'Universit\`a ``La Sapienza'' and Sezione INFN, Roma, Italy.}%
\thanks{$^*$\,Speaker. Address correspondence to Matthew Moulson, e-mail:
\moulson}
}

\maketitle

\begin{abstract}
\boldmath
The P326 experiment at the CERN SPS has been proposed with the purpose of
measuring the branching ratio for the decay $K^+\to\pi^+\nu\bar{\nu}$ to
within $\sim$10\%. 
The photon veto system must provide a rejection factor of \EE{}{8}
for $\pi^0$ decays.
We have explored two designs for the large-angle veto detectors, 
one based on scintillating tiles and the other using scintillating fibers. 
We have constructed a prototype
module based on the fiber solution and evaluated its performance
using low-energy electron beams from the Frascati Beam-Test Facility. 
For comparison, we have also tested a tile prototype constructed for the
CKM experiment, as well as lead-glass modules from the OPAL electromagnetic
barrel calorimeter.
We present results on the linearity, energy resolution, and 
time resolution obtained with the fiber prototype, and compare the 
detection efficiency for electrons obtained with all three instruments. 
\end{abstract}


\section{The P326 Experiment}

\begin{figure}
\centering
\includegraphics[width=0.45\textwidth]{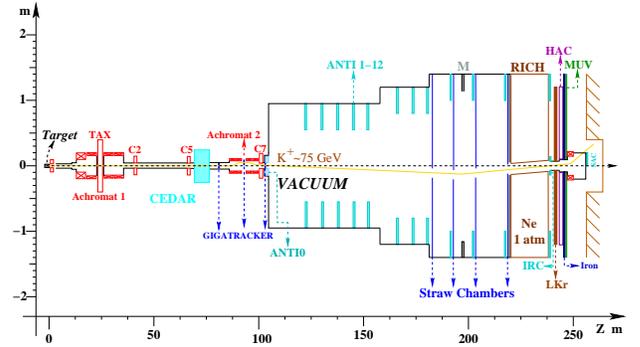}
\caption{The P326 experimental layout.}
\label{fig:expt}
\end{figure}

\IEEEPARstart{T}HE branching ratio (BR) for the decay 
$K^+\to\pi^+\nu\bar{\nu}$ can be 
related to the value of the CKM matrix element $V_{td}$ with minimal 
theoretical uncertainty, providing a sensitive probe of the flavor sector
of the Standard Model.
The measured value of the BR is
\EA{1.47}{1.30}{0.89}{-}{10}
on the basis of three detected events \cite{E949+04:Kpnn}.
P326, an experiment at the CERN SPS, has been proposed with the goal of
detecting $\sim$100 $K^+\to\pi^+\nu\bar{\nu}$ decays with a S/B ratio of
10:1 \cite{P326+05:proposal}. 
The experimental layout is illustrated in \Fig{fig:expt}.

The experiment will make use of a 75 GeV unseparated positive secondary beam.
The total beam rate is 800 MHz, providing $\sim$50 MHz of $K^+$'s.
The decay volume begins 102~m downstream of the production 
target.   
10~MHz of kaon decays are observed in the 120-m long vacuum decay region.
Large-angle photon vetoes are placed at 12 stations along 
the decay region and provide full coverage for decay photons with
$\mbox{8.5~mrad} < \theta < \mbox{50~mrad}$.
The last 35~m of the decay region hosts a dipole 
spectrometer with four straw-tracker stations operated in vacuum. 
The NA48 liquid-krypton 
calorimeter \cite{NA48+07:NIM} is used to veto high-energy photons at
small angle. 
Additional detectors further downstream extend the coverage of
the photon veto system (e.g. for particles traveling 
in the beam pipe).

The experiment must be able to reject background from, e.g., 
$K^+\to\pi^+\pi^0$ decays at the level of \EE{}{12}.
Kinematic cuts on the $K^+$ and $\pi^+$ tracks provide a factor of \EE{}{4}
and ensure 40~GeV of electromagnetic energy in the photon vetoes;
this energy must then be detected with an inefficiency of $\geq \EE{-}{8}$.
For the large-angle photon vetoes, the maximum tolerable detection 
inefficiency for photons with energies as low as 200~MeV is \EE{-}{4}.
In addition, the large-angle vetoes must have good energy and time resolution
and must be compatible with operation in vacuum.

\section{Large-Angle Photon Vetoes}
\label{sec:lav}

The detectors at each veto station are ring shaped. The detectors at the 
first five veto stations have inner radii of 60~cm and outer radii of 96~cm.
Those at the remaining stations have inner and outer radii to match the 
taper of the vacuum chamber; the largest covers
$\mbox{90~cm}<r<\mbox{140~cm}$. 

For the construction of the detectors themselves, two designs
are under consideration.
One consists of a sandwich of lead sheets and scintillating tiles with 
WLS-fiber readout. An assembly of wedge-shaped modules forms the veto station. 
An example of such a detector, using 80 layers of 1-mm thick
lead sheets and 5-mm thick scintillating tiles, was designed for the 
(now canceled) CKM experiment at Fermilab. 
Tests of a prototype at Jefferson Lab showed 
that the inefficiency of the detector for 1.2 GeV electrons was at most 
\EN{3}{-}{6} \cite{RCT04:VVS}.

An alternative solution is based on the design of the KLOE calorimeter 
\cite{KLOE+02:EmC}, and consists of 1-mm diameter scintillating fibers 
sandwiched between 0.5-mm thick lead foils. The fibers are arranged 
orthogonal to the direction of particle incidence and are read out at
both ends. Two U-shaped modules form a veto station. 
This solution offers advantages in terms
of hermeticity, position resolution, and time resolution.
Since a prototype based on the tile design has already been tested, 
we have opted to construct and test a prototype fiber module.
We have obtained the CKM prototype on loan for further testing and 
comparison.

The remainder of this paper describes the construction of the fiber prototype 
and its testing with low-energy electron beams at the Frascati 
Beam-Test Facility (BTF). For the purposes of comparison, we also 
present preliminary results on the detection efficiency for low-energy 
electrons for the CKM tile prototype, and for lead-glass
blocks from the OPAL electromagnetic barrel calorimeter.

\section{Construction of the Fiber Prototype}
\label{sec:cons}

\begin{figure}
\centering
\includegraphics[width=0.25\textwidth]{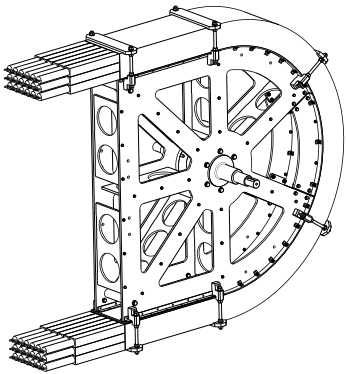}
\caption{Schematic diagram of the fiber prototype attached to 
construction saddle.}
\label{fig:mod}
\end{figure}

One U-shaped module was constructed at Frascati in fall 2006
(\Fig{fig:mod}).
The inner radius (60~cm) and length (310~cm along the inner face) of the 
prototype are identical to the specified values for one of the upstream 
veto stations in the actual experiment.
The prototype has a radial thickness of 12.5~cm, 
corresponding to 35\% of the specified value for one of the upstream
stations. 
This thickness was chosen to reduce prototyping costs; it
should be sufficient for transverse containment of low-energy electron
showers incident half-way between the inner and outer edges of the module.

The materials used in construction were 0.5-mm thick lead foils
from an industrial supplier, cut to $\mbox{350}\times\mbox{25~cm}\up{2}$; 
1-mm diameter Kuraray SCSF-81 scintillating fibers cut to 350~cm in 
length; and Bicron BC-600 optical cement.
Layers of the module were constructed by rolling 1-mm grooves on 1.35-mm 
centers into the lead foils, and gluing scintillating 
fibers into the grooves. 
The 25-cm width of the lead foils determines the longitudinal depth of
the module.
The desired radial thickness was obtained by
stacking 99 layers. 
The ends of the module were then milled
and fitted with
$\mbox{4.2}\times\mbox{4.2~cm}\up{2}$ lucite light guides terminating in
Winston-cone concentrators.
This determines the segmentation of the module in the plane transverse to
the fibers---there are three readout cells in the radial direction and six 
cells in depth.
Light from the fibers is read-out by Hamamatsu R6427 28-mm photomultiplier
tubes (PMTs) coupled to the light guides with optical grease.  

In the region covered by the first four cells in depth, every groove in
the lead is occupied by a scintillating fiber.
In the region covered by the last two cells, the scintillating
fibers in alternating grooves are replaced by lead wire.
This reduces the number of scintillating fibers
by 17\% and increases the thickness
of the module in radiation lengths.
\begin{figure}
\centering
\includegraphics[width=0.2\textwidth]{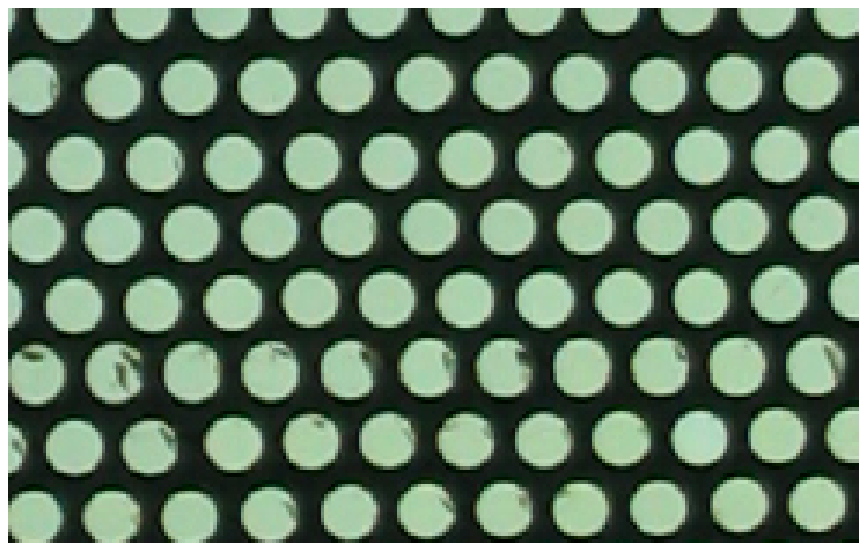}\hspace{1mm}
\includegraphics[width=0.2\textwidth]{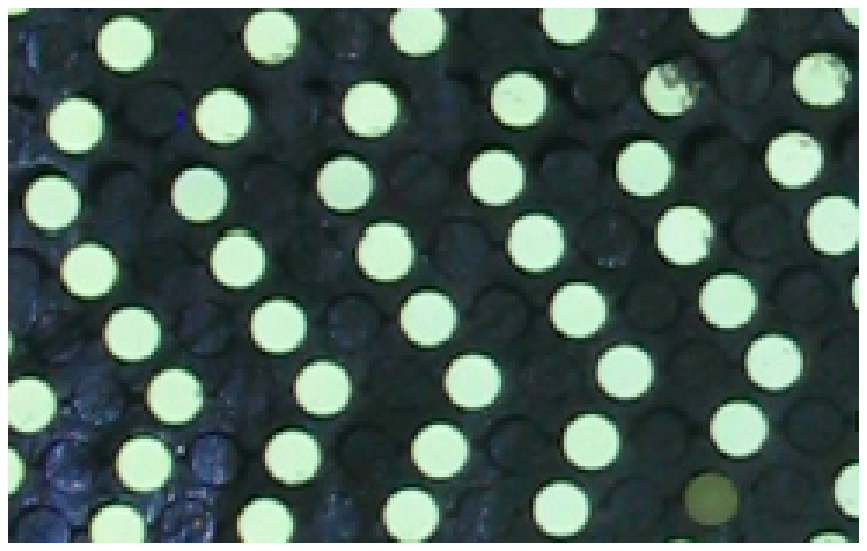}
\caption{Photographs of fiber fill patterns in the (left) first four 
and (right) last two cells in longitudinal depth.}
\label{fig:fill}
\end{figure}
The resulting fill pattern is 
illustrated in \Fig{fig:fill}. 
The first four cells in depth contain lead, fiber, and glue in 
the proportions 42:48:10 by volume, for a thickness of 13$X_0$.
For the last two cells, the proportions are 66:24:10, for a thickness of 
10$X_0$.

For construction of the prototype, we designed the
steel and aluminum support structure, or ``saddle,'' seen in 
\Fig{fig:mod}.
The saddle also provides a convenient support structure for the
completed module, allowing transportation and positioning during beam testing.
However, the module is not 
structurally attached to the saddle. This was intended to allow
experimentation at a later time with module mechanics and support
structures for installation in the actual experiment. 
To avoid complications from nearby material for efficiency 
studies with beam incident near the inner face of the module,
the saddle 
features a removable segment near its midpoint.

The module was constructed as follows.
The lead foils were grooved using a rolling machine built for 
the construction of the KLOE calorimeter \cite{KLOE+02:EmC}.
With the curved surface of the saddle upwards, a foil was draped over 
the saddle, the surface of the foil was painted with optical cement, 
and fibers were laid into the grooves. 
For the 16.8~cm in width corresponding to the first four
readout cells in module depth, every groove is occupied by a fiber.
In this region, the fibers were laid into the grooves \emph{en masse} 
by simply distributing a counted number of fibers over the surface
and smoothing them into place by hand. 
For the last two cells,
fibers and lead wires were individually placed.
The fibers were then painted with optical cement, and the next foil 
was draped on top, providing the bottom surface of the next layer. 
Four to six layers could be completed within the 
2.5-hour pot life of the glue, after which time, a harmonic-steel band 
with threaded tensioning rods at either end was applied and tightened. 
Additional compression was provided by a series of clamps on the saddle.
The module was left to set and cure overnight, and construction proceeded the
following day.

When construction was complete and the module was instrumented
as described above, the responses of the cells to minimum-ionizing
particles were equalized to within 10--20\% by adjusting the PMT 
voltages in successive cosmic-ray runs, 
in which the calorimeter was oriented with the U opening upward, and
scintillator paddles were placed above and below the midpoint of the 
module.

\section{Other Prototypes}

As described in \Sec{sec:lav}, the prototype lead/scintillating-tile
photon veto detector constructed for the CKM experiment was obtained 
on loan from Fermilab for testing and comparison. This detector 
is fully described in \Ref{RCT04:VVS}.

In addition, 3800 modules from the central part of the OPAL 
electromagnetic calorimeter barrel \cite{OPAL+91:NIM}
have recently become available 
for use in P326. These modules consist of blocks of
SF57 lead glass with an asymmetric, truncated square-pyramid shape.
The front and rear faces of the blocks measure about 
$\mbox{10}\times\mbox{10~cm}\up{2}$ and
$\mbox{11}\times\mbox{11~cm}\up{2}$, respectively; the blocks are 
37~cm long.
The modules are read out at the back side by Hamamatsu R2238 (76-mm diameter) 
PMTs, coupled via 4-cm cylindrical light guides of SF57.
There are obvious practical advantages to basing the P326 large-angle
photon veto system on existing hardware; the collaboration is actively
seeking to understand whether these Cerenkov radiators are suitable.
For most of our beam tests, we used a tower of four lead-glass blocks, with
the beam centrally incident on the side face of the 
first module. In this configuration, the stack was $\sim$40~cm
(27$X_0$) deep.

\section{The Frascati Beam-Test Facility}

The Frascati BTF \cite{M+03:BTF} is an electron transfer line leading 
off of the DA$\Phi$NE linac.
The linac accelerates $e^+$'s and $e^-$'s to maximum energies of 550 and
800~MeV, respectively, producing 10-ns pulses with a repetition rate of 50~Hz.
Momentum-selection magnets, attenuating targets, and collimation slits 
upstream 
of the experimental area can be used to produce test beams in the BTF 
hall with energies from
$\sim$100 to 750~MeV with a 1\% energy-selection resolution 
and mean multiplicities from $<$1 to \EE{}{9} per pulse.

The last magnet on the BTF line is a 45$^\circ$ dipole with a hole in the 
yoke allowing extraction of a photon beam through an uncurved extension of the 
vacuum chamber.
The apparatus for producing a tagged photon beam was developed 
for testing the AGILE satellite gamma-ray telescope \cite{H+05:AGILE}. 
A silicon microstrip beam tracker doubles as an active
bremsstrahlung target upstream of the final dipole; 
silicon trackers inside the dipole gap register the 
trajectory of the electron after radiation, tagging the bremsstrahlung 
photon.
While the tagged photon beam has been used with some success for
energy calibration, e.g., of the AGILE satellite,
at present, background levels in the photon beam
are prohibitive for sensitive efficiency measurements.
This background consists of photons from showering on upstream
beam elements by particles from the attenuating target.  
Work is underway to improve the shielding around the attenuating target.
In the meantime, we have used the electron beam to test the 
prototypes.

\section{Readout and data acquisition}

All prototypes were read out using the BTF front-end electronics and 
data acquisition (DAQ) system. For the fiber and tile prototypes, 
the PMT anode signals were passively split to obtain both 
charge and time measurements. For the lead-glass detectors,
the signals were not split and only charge information was read out.
CAEN V792 charge-to-digital converters (QDCs) were used for the charge 
measurements.
CAEN V814 low-threshold discriminators and V775 time-to-digital converters
(TDCs) were used for the time measurements.
A signal from the linac gun provided 
QDC gates and TDC starts, as well as the DAQ trigger.

The 12-bit QDCs used reached full scale at 400~pC.
A minimum-ionizing particle passing through the fiber prototype
deposits 30~MeV per cell. 
For efficiency studies, we desired that this correspond to about 
200 QDC counts. 
The HV settings obtained from calibration
with cosmic rays as described in \Sec{sec:cons}
then gave typical tube gains of \EN{1}{}{7}.

\section{Beam tagging system}
\label{sec:tag}

The telescope of scintillation counters used to tag single-electron events
\begin{figure}
\centering
\includegraphics[width=0.45\textwidth]{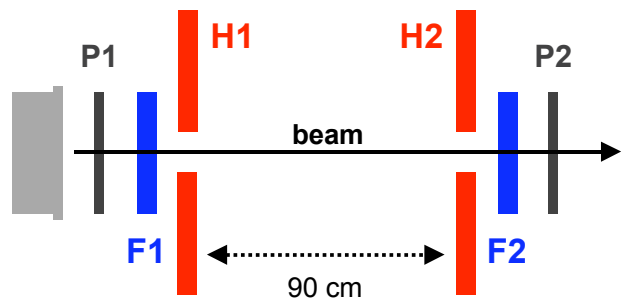}
\caption{Schematic diagram of the beam tagging system, comprising
two paddles for single-electron event selection (F1, F2), two 
hole counters for trajectory definition (H1, H2), and two beam-profile 
monitors for alignment (P1, P2).}
\label{fig:tag}
\end{figure}
is schematically illustrated in \Fig{fig:tag}.
From upstream to downstream, the following trigger counters,
all made from 10-mm thick plastic scintillator, were used:
\begin{itemize}
\item[F1] a paddle of area $\mbox{60}\times\mbox{85}$~mm\up{2},
positioned a few centimeters downstream of the 
beamline exit window;
\item[H1] a paddle of area $\mbox{200}\times\mbox{130}$~mm\up{2}
with a 14-mm diameter hole in the center, 
positioned $\sim$10~mm downstream of F1;
\item[H2] a paddle of area $\mbox{330}\times\mbox{100}$~mm\up{2}
with a 14-mm diameter hole in the center, 
positioned 90~cm downstream of H1.
\item[F2] a paddle of area $\mbox{60}\times\mbox{85}$~mm\up{2},
positioned $\sim$10~mm downstream of H2 and as little as 10~mm
upstream of the prototype to be tested.
\end{itemize}
The tagging criterion for single-electron events used in the 
efficiency studies was
$\mbox{F1}\cdot\overline{\mbox{H1}}\cdot\overline{\mbox{H2}}\cdot\mbox{F2}$,
where F1 and F2 refer to charge signals on the paddle counters consistent
with passage of a single electron, and $\overline{\mbox{H1}}$ and
$\overline{\mbox{H2}}$ refer to null signals on the hole counters.
Acceptable beam trajectories were thus defined by the
two 14-mm diameter holes separated by 90~cm.
The use of paddle/hole combinations rather than 
horizontal/vertical fingers was intended to reduce the amount of material in
the beam.
The fact that no material occupied the space between the 
hole counters was intended to facilitate alignment.
The thickness of the paddles was chosen to allow 
efficient identification of events with exactly
one electron in the paddles within the 10-ns linac pulse.
The large dimensions of the hole counters served to help reject events 
with stray beam particles present.  
The use of a paddle (rather than a hole) as the last counter was intended 
to reduce mistags by providing a positive signal for beam particles just 
before entry into the prototype.

The mistag probability was monitored by taking occasional runs with
the last dipole in the BTF beamline switched off, so that the beam was
not directed towards the tagger or the prototypes. We did not find
any tags in more than 1 million events collected in this configuration,
corresponding to a false-tag rate of $< \EN{2}{-}{6}$ at 90\% CL. 
Based on our evaluation of the efficiencies 
for the F1 and F2 counters singly, we expect the contribution from false
tags to be insignificant for the purposes of the efficiency measurements.
In all cases, we quote efficiencies assuming no contribution from 
false tags. This assumption is conservative; if there are false tags,
they artificially increase the inefficiency. 
The energy spectrum from the fiber prototype for fully-tagged events
shows that, at most, multiple-electron events are present at the percent
level (\Fig{fig:eff}), and thus have a negligible effect on the normalization
for the efficiency measurements.  
 
The tagging system was mounted on a rigid support structure allowing 
fine and reproducible positioning of all counters in the horizontal 
and vertical coordinates. To facilitate alignment, the beam position in 
the bend plane was measured using the BTF beam-profile meters, which
were mounted just 
upstream and downstream of the tagger (P1 and P2 in \Fig{fig:tag}). 
Each profile meter is a
one-dimensional, 16-channel close-packed array of 1-mm scintillating 
fibers read out by a multianode PMT, with each channel consisting of a 
group of fibers three across by four deep.

\section{Data Collection}

The fiber prototype was first tested at the BTF
during two-week runs in March and April 2007. During the April run,
the tile prototype was also tested. These runs served primarily
to debug the prototypes and optimize the tagging system.
\begin{table}
\renewcommand{\arraystretch}{1.3}
\caption{Statistics Collected During the Summer 2007 BTF Run}
\label{tab:stat}
\centering
\begin{tabular}{ccccc}
\hline\hline
& Beam energy [MeV] & \Pmul\ [\%] & \etag\ [\%] & Tagged events\\
\hline
\multicolumn{2}{l}{Fiber prototype (KLOE)} &  & \\
& 203 & 31.3 & 2.5 & 70k \\
& 350 & 33.0 & 9.2 & 210k \\
& 483 & 33.3 & 14.4 & 370k \\
\hline
\multicolumn{2}{l}{Tile prototype (CKM)} &  & \\
& 203 & 29.5 & 3.7 & 65k \\
& 350 & 31.8 & 8.8 & 220k \\
& 483 & 29.0 & 17.6 & 370k \\
\hline
\multicolumn{2}{l}{Lead glass (OPAL)} &  & \\
& 203 & 30.2 & 3.9 & 25k \\
& 483 & 26.0 & 17.1 & 90k \\
\hline\hline
\end{tabular}
\end{table}

The data analyzed for this report were collected during a 25-day run
during June-July 2007. For the fiber and tile prototypes, data were
taken at beam energies of 203, 350, and 483~MeV. For the lead-glass
detectors, data were taken at beam energies of 203 and 483 MeV.
\Tab{tab:stat} summarizes the data collected. For each instrument 
and point in beam energy, the total number of fully-tagged single-electron
events is given, together with \Pmul, the probability for having 
an event of multiplicity one in the prototype 
(by Poisson statistics, $\Pmul = \mbox{36}$\%
in the best case), and \etag, the fraction of
single-electron events in the prototype
passing the full tagging criterion.
There is a marked effect from multiple scattering in the tagger and in air:
the tagging efficiency \etag\ is significantly
decreased at low energy.
  
In tests of the tile prototype at Jefferson Lab, data were taken at 
beam energies of 500, 800, and 1200~MeV. 
Our tests thus extend to significantly lower 
beam energies the experimental knowledge of the detection efficiency 
for this prototype.

In addition to the data samples listed in \Tab{tab:stat}, for each detector,
smaller samples were also collected in a variety of configurations with 
the beam incident at different points and/or at different angles.

\section{Results Obtained with the Fiber Prototype}

\subsection{Energy Reconstruction}
\label{sec:erec}

We obtain separate energy measurements from the set of PMTs on 
each side of the prototype (sides A and B). We first subtract the
mean noise level from the QDC measurements for each cell.
The noise arises from diffuse background in the BTF hall; its 
mean level is determined from events with no activity in the
tagger, and is typically larger than the sum of the hardware pedestals
by an amount corresponding to a few MeV for the whole detector. 

For each side, 
we take the energy measurement to be the gain-calibrated sum of the signals 
from all cells for which the uncalibrated QDC measurements are greater 
than the hardware pedestal 
by more than 3$\sigma$ (typically less than 10 counts, or $\sim$1.5~MeV).
For the combined energy measurement from both sides, if there
are signals above the 3$\sigma$ threshold from both PMTs,
the energy measurement for the cell is the average of the measurements
from each side. If instead one PMT gives a signal above threshold and 
the other does not, the energy measurement for the cell is equal 
to the measurement from the side above threshold. 

For some runs with $E\dn{beam} = \mbox{350}$ and 500~MeV, a few of the QDC 
channels digitizing signals from side B of the prototype refused to 
register any counts above pedestal unless the integrated PMT signals 
exceeded the normal level of the hardware pedestal by $\sim$100 counts.
This led to a loss of part of the signal from side B at low energies.
When the energy measurements from both sides are combined, the 
use of the algorithm described above helps to recover the lost signal.

\subsection{Linearity and Energy Resolution}

Although seemingly a basic test of the prototype performance, the
linearity of response is difficult to measure precisely with our setup.
This is mainly because run-to-run fluctuations in the energy scale are
observed at the $\sim$5\% level. Several factors may contribute to such
drifts, including limited reproducibility of the beam energy due to
hysteresis in the BTF dipoles and possible time- (or temperature-) 
dependent drifts in HV power supply voltages or QDC gains.
With additional effort during data taking, it should be possible to 
maintain better stability of the energy scale. In any event, for the 
energy resolution and efficiency measurements, we calibrate to a reference
value of the energy for the single-electron peak, so these small 
drifts do not pose a problem. When testing the linearity, however, this
calibration procedure cannot be applied at more than one energy point.
\begin{figure}
\centering
\includegraphics[width=0.45\textwidth]{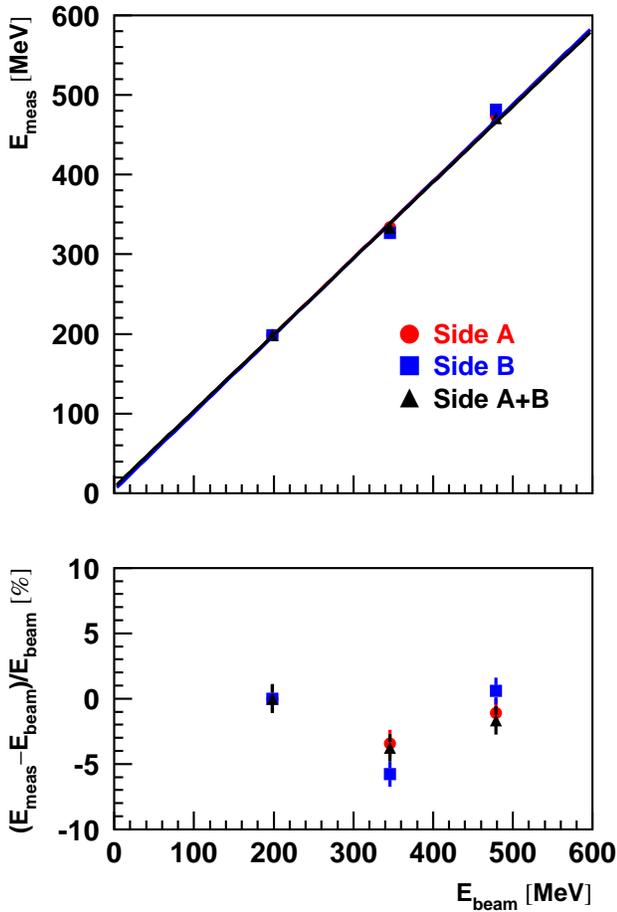}
\caption{Response linearity for the fiber prototype. 
Top: $E\dn{meas}$ vs.\ $E\dn{beam}$ for $E\dn{beam} = \mbox{203}$, 350, and 
483~MeV. Bottom: $(E\dn{meas} - E\dn{beam})/E\dn{beam}$ for each point.}
\label{fig:lin}
\end{figure}
In \Fig{fig:lin}, we plot the measured mean value of the energy of the 
single-electron peak, $E\dn{meas}$, as a function
of the beam energy, $E\dn{beam}$, where the energy scale has been fixed 
using the point at $E\dn{beam} = \mbox{203 MeV}$.
$E\dn{meas}$ is obtained from Gaussian fits to the single-electron peak
over an interval of about $\pm$1.5$\sigma$ about the peak.  
The lower panel of the figure
shows the fractional deviation of $E\dn{meas}$ from $E\dn{beam}$. 
Such deviations are present at the level of $\sim$5\%, i.e., at the 
level of precision with which the energy scale is known.
(The errors on the plotted points include only the statistical measurement
errors, plus a 1\% systematic error corresponding to the BTF energy-selection
resolution.)

We conclude that the response linearity is basically satisfactory.
We do note that, in the actual experiment, multiple-range QDCs or some
other readout scheme with extended dynamic range will be necessary, 
as full-scale
is reached on the front cells for multiple-electron events in which the 
total energy deposit is $\sim$2~GeV.  
\begin{figure}
\centering
\includegraphics[width=0.45\textwidth]{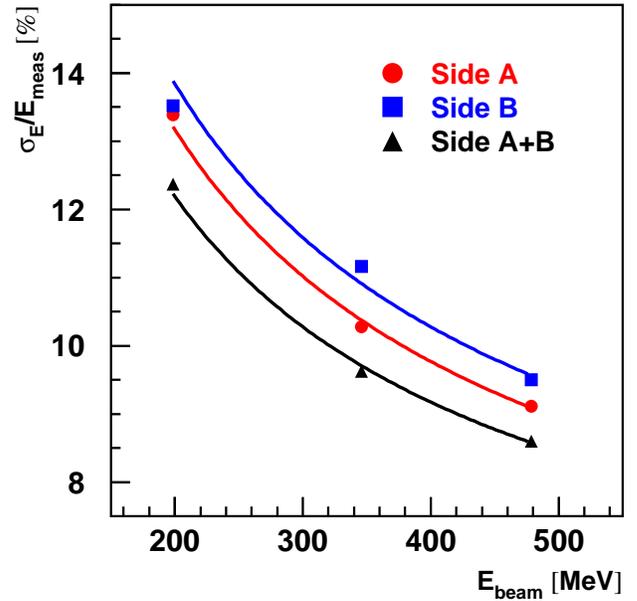}
\caption{Energy resolution for the fiber prototype at 
$E\dn{beam} = \mbox{203}$, 350, and 483~MeV.}
\label{fig:res}
\end{figure}

To obtain the energy resolution, the Gaussian fits to the 
single-electron peak are performed again after the run-by-run 
energy scale calibration is applied.
In \Fig{fig:res}, we plot the relative energy resolution, 
$\sigma_E/E\dn{meas}$, as a function of $E\dn{beam}$, for the measurements
from each side of the prototype and for the combined measurement. 
The inferior performance of side B at $E\dn{beam} = \mbox{350}$ and 500~MeV
is due to the loss of signal described in \Sec{sec:erec}.
The best performance is obtained by combining information from both sides.
The curves in \Fig{fig:res} show the results of fits to the form
\begin{displaymath}
\frac{\sigma_E}{E} = \frac{p\dn{1}}{\surd{E}\:\mbox{(GeV)}} \oplus p\dn{2}.
\end{displaymath}
Using the information from both sides of the
prototype, we find $p\dn{1}$ = 5.1\% and $p\dn{2}$ = 4.4\%.
The coefficient obtained for the stochastic term ($p\dn{1}$) is 
in reasonable agreement with expectation from our preliminary Monte Carlo 
studies and from experience with the KLOE calorimeter.

\subsection{Time resolution}

In principle, the arrival time of a particle and its impact position 
along the length of the fibers would be obtained from the sum (average) 
and difference of the time measurements from the two sides of a cell. 
However, for the tests described here, the beam was incident at 
the midpoint of the fibers; we therefore have independent time measurements 
from each side of each cell. 
The time measurements for sides A and B, $t\dn{A}$ and $t\dn{B}$, 
and the combined time measurement $t\dn{A+B}$,
are taken to be the energy-weighted averages of the time measurements for
the corresponding group of cells. The event time reference is provided 
by the tagging system: $t\dn{tag} \equiv (t\dn{F1} + t\dn{F2})/\mbox{2}$, 
where F1 and F2 are the trigger paddles described in \Sec{sec:tag}. 
  
Slewing corrections and time offsets for each cell are obtained by 
fitting the time vs.\ QDC distributions with the form
$t - t\dn{tag} \propto (\mbox{ln}\:Q_0/Q)^\alpha + t_0$, where
$Q$ and $t$ are the QDC and time measurements, $t_0$ is the time offset
for the cell, and $\alpha$ is positive.
Slewing corrections are also necessary for $t\dn{F1}$ and $t\dn{F2}$, 
so an iterative procedure is applied.

Once all slewing corrections have been obtained, we form the distributions
of the differences
$t\dn{A} - t\dn{tag}$,
$t\dn{B} - t\dn{tag}$,
$t\dn{A} - t\dn{B}$, and
$t\dn{F1} - t\dn{F2}$; fit with Gaussians; and from the four widths 
obtain $\sigma\dn{A}$, $\sigma\dn{B}$, $\sigma\dn{tag}$, and $\sigma\dn{AB}$,
where this latter quantity accounts for common-mode fluctuations in the 
time measurements from the two sides 
($\sigma\dn{AB}\up{2} = 2\,\mbox{cov}(t\dn{A},t\dn{B})$).
The time resolution of the tagging system is found to be
$\sigma\dn{tag} \approx \mbox{147 ps}$ and stable for points with different
$E\dn{beam}$.  
We obtain the resolution on the combined time measurement for the two sides
from the width of the distribution $t\dn{A+B} - t\dn{tag}$, with 
$\sigma\dn{tag}$ subtracted in quadrature.
\begin{figure}
\centering
\includegraphics[width=0.45\textwidth]{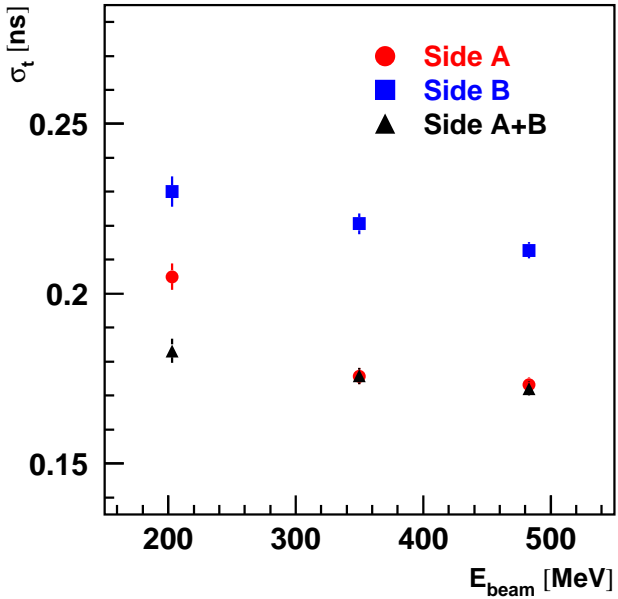}
\caption{Time resolution for the fiber prototype at 
$E\dn{beam} = \mbox{203}$, 350, and~483 MeV (preliminary)}
\label{fig:time}
\end{figure}

Our results on the time resolution are still preliminary. They are 
plotted in \Fig{fig:time} as a function of $E\dn{beam}$.
Again, the resolution is better on side A than it is on side B. 
For the point at 483 MeV, the resolution for the combined measurement
is $\sigma\dn{A+B} = \mbox{172 ps}$, of which 158~ps is due to the common-mode
fluctuation in the time measurements from each side. 
We do not yet fully understand the origin of this large contribution.

\subsection{Efficiency}
\begin{figure*}
\centering
\includegraphics[width=0.85\textwidth]{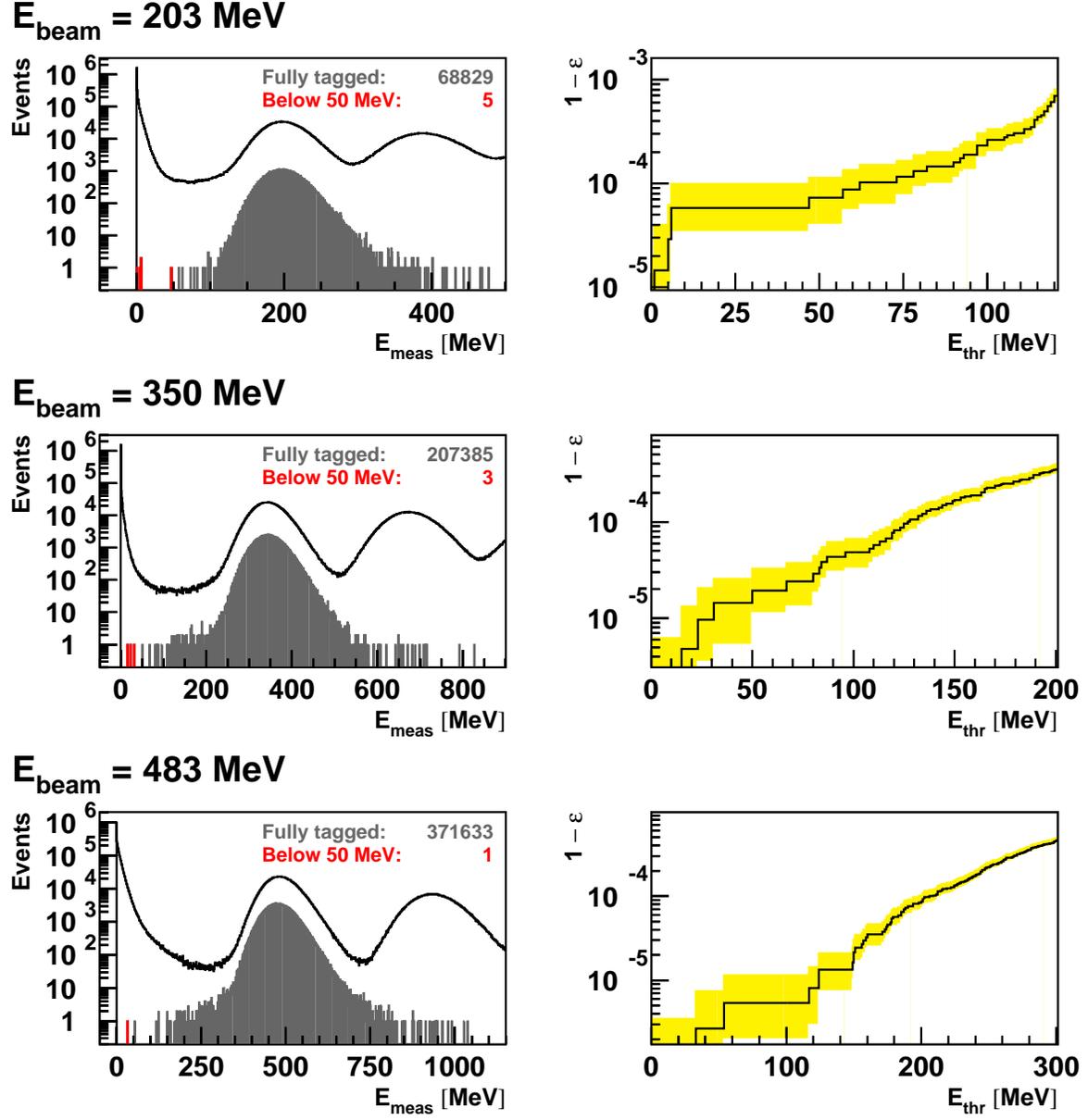}
\caption{Measurements of detection efficiency for electrons with 
$E\dn{beam} = \mbox{203}$, 350, and 483~MeV. 
Left: Measured energy distributions 
for all events (open histograms) and for fully tagged events 
(shaded histograms). Right: Inefficiency as a function of $E\dn{thr}$.}  
\label{fig:eff}
\end{figure*}

Our measurements of the detection efficiency are summarized in 
\Fig{fig:eff}.
For each beam energy, the panel on the left shows the
energy distribution for all collected events (open histogram) and for 
fully-tagged events (shaded) histogram.
The one- and two-electron peaks are clearly visible in the distribution
for all events; application of the tagging criterion reduces the 
contribution from multiple-electron events to a level negligible for our 
purposes.

We consider a fully-tagged single-electron event to be 
undetected if the measured energy is below a threshold value of
$E\dn{thr} = \mbox{50 MeV}$. 
At $E\dn{beam} = \mbox{203 MeV}$, we find five such events out of 68$\,$829
total tagged events; at $E\dn{beam} = \mbox{350 MeV}$, we find 
three out of 207$\,$385; and at 483~MeV, we find one out of 371$\,$633.
We thus quote inefficiencies of 
\EA{7.3}{4.1}{3.3}{-}{5},
\EA{1.4}{1.1}{0.9}{-}{5}, and
\EA{2.7}{4.7}{1.7}{-}{6}, respectively, where the asymmetric uncertainties
represent 68.27\% unified confidence intervals \cite{FC98:CL}.
We assume that no undetected events are due to false tags. 

The choice of threshold $E\dn{thr} = \mbox{50 MeV}$
is reasonable but arbitrary.
For each beam energy, we obtain the inefficiency as a function of 
threshold from the normalized cumulative energy distribution 
for fully-tagged events.
The results are presented in the right panels of \Fig{fig:eff}.
One again, the shaded bands indicate 68.27\% unified confidence 
intervals, and we assume that there are no false tags. 
For $E\dn{beam} = \mbox{203 MeV}$, the inefficiency remains
at the level of a few per mil even for thresholds as high as 100~MeV.

\subsection{Comparison with Other Prototypes}

The analysis of the data from the tile and lead-glass detectors is not
yet complete, in particular with respect to the final, run-dependent
energy calibrations. 
\begin{table*}
\renewcommand{\arraystretch}{1.3}
\caption{Comparison of Detection Efficiencies for Three Prototypes}
\label{tab:comp}
\centering
\begin{tabular}{ccccc}
\hline\hline
& Beam energy [MeV] & Tagged events & Tagged, $E\dn{meas} < \mbox{50 MeV}$ 
& $\mbox{1} - \varepsilon$ \\
\hline
\multicolumn{5}{l}{Fiber prototype (KLOE)} \\
& 203 &  68$\,$829 & 5 & \EA{7.3}{4.1}{3.3}{-}{5} \\
& 350 & 207$\,$385 & 3 & \EA{1.4}{1.1}{0.9}{-}{5} \\
& 483 & 371$\,$633 & 1 & \EA{2.7}{4.7}{1.7}{-}{6} \\
\hline
\multicolumn{5}{l}{Tile prototype (CKM) - Preliminary} \\
& 203 &  65$\,$165 & 2 & \EA{3.1}{3.5}{1.9}{-}{5} \\
& 350 & 221$\,$162 & 3 & \EA{1.4}{1.0}{0.9}{-}{5} \\
& 483 & 192$\,$412 & 1 & \EA{5.2}{9.1}{3.3}{-}{6} \\
\hline
\multicolumn{5}{l}{Lead glass (OPAL) - Preliminary} \\
& 203 &  25$\,$069 & 3 & \EA{1.2}{0.9}{0.8}{-}{4} \\
& 483 &  91$\,$511 & 1 & \EA{1.1}{1.9}{0.7}{-}{5} \\
\hline\hline
\end{tabular}
\end{table*}
\begin{figure}
\centering
\includegraphics[width=0.45\textwidth]{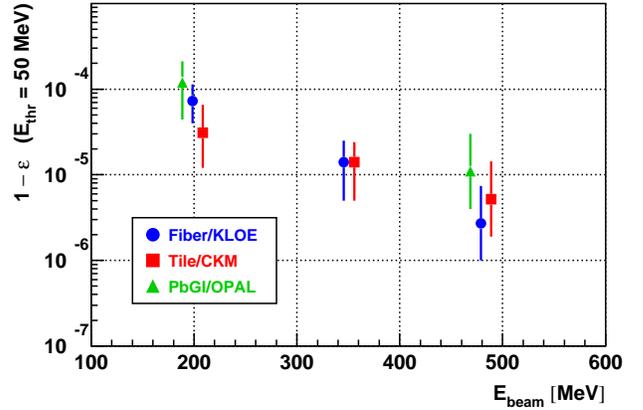}
\caption{Comparison of detection efficiencies for each of the three 
prototypes, for 203, 350, and 483 MeV electrons with 
$E\dn{thr} = \mbox{50 MeV}$.
Results obtained with the CKM tile prototype and OPAL lead-glass modules are
preliminary.}
\label{fig:comp}
\end{figure}
Nevertheless, we believe that our preliminary 
results on the detection efficiencies for these prototypes are
sufficiently stable to provide meaningful comparison with
the results obtained with the fiber prototype. The results obtained 
for the inefficiency with $E\dn{thr} = \mbox{50 MeV}$ 
for all three prototypes are 
summarized in \Tab{tab:comp} and plotted in \Fig{fig:comp}.
The efficiency for detection of low-energy electrons is seen to be 
similar for all three technologies tested.

\section{Summary and Outlook}

The large-angle photon veto detectors for the P326 experiment must have
inefficiencies of less than \EE{-}{4} for the detection of photons with
energies as low as 200~MeV. We have constructed a lead/scintillating-fiber 
prototype detector based on the KLOE calorimeter and tested it with 
electrons at the Frascati BTF. The performance of the prototype is in
agreement with expectation; in particular, we obtain an energy 
resolution of
$\sigma_E/E = \mbox{5.1\%}/\surd E\:\mbox{(GeV)} \oplus \mbox{4.4\%}$
and an inefficiency for the detection of 203~MeV electrons of 
\EA{7.3}{4.1}{3.3}{-}{5}.

A preliminary analysis of data from the CKM tile prototype and from
lead-glass modules from the OPAL barrel electromagnetic calorimeter
suggests that all three detectors have similar detection efficiencies for
electrons. 
However, the detection efficiency for photons may be worse,
whether because of punch-through, or because of a high intrinsic 
inefficiency for the detection of photonuclear interactions
\cite{A+99:photonucl,A+05:photonucl}.

\enlargethispage{-160mm}
Since there is a significant practical advantage to basing 
the P326 photon vetoes on existing hardware, our focus for the near-term
future will be on investigation of the OPAL lead-glass modules
as an appropriate technology on which to base the P326 low-energy photon
vetoes.

\section*{Acknowledgment}

We would like to thank B.~Buonomo and G.~Mazzitelli (Frascati)
for their assistance with BTF operations during our data-taking periods.
We would also like to thank S.~Cerioni and B.~Dulach (Frascati) 
for the designs of the construction saddle and the mechanical support
for the tagging system, as well as for their assistance
with various construction issues.

\bibliographystyle{IEEEtran}
\bibliography{hawaii_proc}

\end{document}